\documentclass[%
showpacs,
nofootinbib,
amsmath,amssymb,
prd,
twocolumn,
superscriptaddress]{revtex4-2}

\usepackage{graphicx}
\usepackage[all,arc,knot,matrix,color]{xy}
\usepackage{braket}
\usepackage{natbib}
\usepackage{hyperref}
\usepackage[mathcal]{euler}
\usepackage{amsmath, amssymb}

\begin{document}

\title{A no-go study on gravitational helicity in connection variables}

\author{Xiao-Kan Guo}
 \email{kankuohsiao@whu.edu.cn}
\affiliation{School of Mathematics \& Physics, Yancheng Institute of Technology, Jiangsu 224051, China}
\author{Shupeng Song}
 \email{songsp@ustb.edu.cn}
\affiliation{Basic Experimental Center for Natural Science, University of Science and Technology Beijing, Beijing 100083, China}
\date{\today}

\begin{abstract}
We study the gravitational helicity using the covariant phase space method. Starting from the
covariant phase space of general relativity expressed in terms of connection variables, we
construct the symplectic form and identify a duality transformation based on the internal Hodge
dual. In the complex self-dual Ashtekar variables the duality becomes a simple $U(1)$ phase
rotation. We show, however, that this rotation is not
a symmetry.
We also show that the expression $\int_\Sigma\Sigma^{+AB}\wedge A^+_{AB}$
inspired by the electromagnetic helicity is neither a Noether charge nor a
moment map.
The  duality of general relativity under a
rotation of the curvature two-form is a symmetry of the equations of
motion but not of the Lagrangian, and carries no Noether charge. An explicit computation shows
that the mass and the NUT charge are not symplectically conjugate, so even in the reduced phase
space of stationary solutions this duality has no moment map. This study therefore provides counter examples to the existence of a full-theory gravitational helicity in the covariant phase space approach.
\end{abstract}

\maketitle

\section{Introduction}
The optical helicity of electromagnetic fields is known to be the conserved quantity associated
with the duality symmetry of the vacuum Maxwell equations \cite{Cal65,DT76}. The conservation
of helicity under duality transformation can also be understood by embedding Maxwell's
theory into a duality-symmetric theory \cite{Ran92,CB12,BBN13}.
It is found that the optical helicity can be expressed as the difference of two Chern-Simons
terms; the appearance of topological numbers agrees with the physical interpretation of the
optical helicity as the difference of left and right handed photons \cite{AS96}.

Recently in \cite{EDHZ16}, a symplectic approach to derive the electromagnetic helicity is
presented, which is based on the covariant phase space method \cite{CW87}. In this approach,
the optical helicity in Chern-Simons form is recovered by the moment map of the duality
transformation. The use of the covariant phase space method greatly simplifies the derivation.

Based on the formal similarity between Maxwell's electromagnetism and linearized general
relativity, the similar story of helicity has been told in linearized gravity
\cite{Bar14,AAB21,Tot22}. However, this kind of duality (and hence
helicity) for linearized gravity cannot be generalized to full nonlinear general
relativity if one directly uses the spacetime Hodge duality to define the electromagnetic
duality \cite{DS05}. The point is that the linearized gravitational fields are actually
in the tangent space of the covariant phase space of the full nonlinear theory, so directly
working on the ``base space'' fails to retain the duality. This point of view is corroborated
by the recent work \cite{Kol22} where the tangent space Hodge dual, i.e. the internal Hodge dual,
is exploited to define the duality for general relativity written in terms of tetrad variables.

An interesting thing is that, in analogy to the electromagnetic helicity appearing as the
Chern-Simons term, the corresponding topological term resulting from the duality in general
relativity is expected to be the Nieh-Yan term \cite{NY82}. This possibility has been discussed in
\cite{Kol22}, and more recently in \cite{LZ26} it has been explicitly realized that the gravitational helicity flux is related to the Nieh-Yan term reduced on the null hypersurfaces.

The Nieh-Yan term, when included in the gravity action, is important for
introducing the Barbero-Immirzi parameter $\gamma$ of loop gravity \cite{Bar95,Imm97,Holst,Mercuri09,DKS09, RW12,Bianchi24}. Therefore, we study in this paper the gravitational helicity in the connection formulation of general relativity (or loop gravity) by using the covariant phase space method. Our result shows that, however,   the
$U(1)$ symmetry that exists is a rotation of the {\it curvature}, not of the connection. This corroborates the study by Kol \cite{Kol22,KolYau23}, but still faces the argument that
that such a symmetry is broken by interactions \cite{Monteiro23}.

We begin in the next section by recalling the covariant phase space approach to the
electromagnetic helicity. Then in Sec.~\ref{S3} we develop the analogous
construction for general relativity in the complex self-dual Ashtekar variables, where the
internal duality becomes a $U(1)$ phase rotation $A^+\to e^{i\theta}A^+$ of the connection, and we
show that the candidate charge suggested by this rotation is not a valid helicity. In
Sec.~\ref{sec:KTN} we evaluate this candidate charge on the Kerr-NUT solution as an example and
find that it vanishes. In Sec.~\ref{sec:curv} we recall the correct duality, a rotation of the
curvature. Sec.~\ref{IV} contains the conclusion and some discussions.

\section{Electromagnetic helicity}
Let us start with the action for the vacuum electromagnetic fields
\begin{equation}\label{1}
S=-\frac{1}{4}\int_\mathsf{M} d^4x\,F_{\mu\nu}F^{\mu\nu}
\end{equation}
on Minkowski spacetime $\mathsf{M}$. The action is invariant under the duality transformation
\begin{equation}\label{2}
F\mapsto\hat{F}=F\cos\theta +(*F)\sin\theta,\quad\theta\in\mathbb{R},
\end{equation}
up to a surface term, where $F=dA$ and $*F$ is its spacetime Hodge dual.
To construct the covariant phase space, consider a Cauchy surface $\Sigma$.
The variation of the action on a region $\mathsf{M}'=[t_0,t_1]\times\Sigma$ yields the symplectic current,
whose integral over $\Sigma$ defines the presymplectic 2-form
\begin{equation}\label{7}
\omega(\delta A,\delta' A)=\int_\Sigma\big(\delta A\wedge*(\delta'F)-\delta'A\wedge*(\delta F)\big).
\end{equation}
Quotienting by gauge transformations $A\to A+d\varphi$ gives the physical phase space
$\mathcal{M}$ on which $\omega$ is symplectic.

If we introduce a dual gauge potential $C$ such that $*F=dC$, the duality transformation
\eqref{2} can be implemented at the level of connections as a rotation between $A$ and $C$.
The infinitesimal generator $\delta_\epsilon A=\epsilon C$, $\delta_\epsilon C=-\epsilon A$
is Hamiltonian, and its moment map gives the optical helicity
\begin{equation}
J=-\frac{1}{2}\int_\Sigma (A\wedge dA+C\wedge dC),
\end{equation}
which equals the difference $N_L-N_R$ of left and right handed photons \cite{AS96,EDHZ16}.

Two ingredients are essential in this construction: the duality is a {\it symmetry of the
action} (it changes the Lagrangian by a total derivative $F\wedge F=d(A\wedge F)$), and it is a
{\it canonical transformation} of the phase space (it preserves $\omega$). This symplectic
approach to electromagnetic helicity motivates us to study the corresponding structure in the
covariant phase space of loop gravity. As we will see, {\it both} ingredients fail for the
connection-formulation duality.
\section{Gravitational helicity in connection variables}\label{S3}
In this section we start with the covariant phase space formulation for loop gravity. Then we
identify the duality transformation based on the internal Hodge dual and examine whether it
defines a helicity. To make the duality symmetry completely transparent, we work with the
{\it complex self-dual} Ashtekar variables \cite{Ash86}.

\subsection{Covariant phase space for self-dual loop gravity}
The covariant phase space for loop gravity has been investigated, for example, in
\cite{DH97,LSZ98}. This provides the natural arena for discussing
duality symmetries in the full theory.

Let $(M,g)$ be a four-dimensional spacetime with a foliation $M=\mathbb{R}\times\Sigma$.
The complex self-dual Ashtekar connection $A^+_{AB}$ is defined as the projection of the spin
connection $\omega_{AB}$ onto the self-dual subspace of the Lorentz Lie algebra:
\begin{equation}
A^+_{AB} = \frac{1}{2}\big( \omega_{AB} - i \star\omega_{AB} \big),\qquad
\star A^+_{AB} = i A^+_{AB},\label{5}
\end{equation}
where $(\star\omega)_{AB} = \frac{1}{2}\epsilon_{ABCD}\,\omega^{CD}$ is the internal Hodge dual.
The conjugate momentum is the self-dual 2-form
\begin{equation}
\Sigma^+_{AB} = \frac{1}{2}\big( \Sigma_{AB} - i \star\Sigma_{AB} \big),\qquad
\Sigma_{AB} = \frac{1}{2}\epsilon_{ABCD}\, e^C\wedge e^D,\label{6}
\end{equation}
constructed from the tetrad $e^A$. The action for general relativity in these variables takes the
form
\begin{equation}
S[e,A^+] = \int_M \Sigma^{+ AB} \wedge F^+_{AB},
\label{eq:S_selfdual}
\end{equation}
where $F^+_{AB} = dA^+_{AB} + A^+_A{}^C \wedge A^+_{CB}$ is the curvature of the self-dual connection.

Varying the action with respect to $A^+$ and $e$ (or equivalently $\Sigma^+$) yields the field
equations. Indeed, consider a general variation of the action \eqref{eq:S_selfdual}:
\begin{equation}
\delta S = \int_M \Big( \delta\Sigma^{+ AB} \wedge F^+_{AB} + \Sigma^{+ AB} \wedge \delta F^+_{AB} \Big).
\end{equation}
The variation of the curvature is $\delta F^+_{AB} = d_{A^+}(\delta A^+_{AB})$, where
$d_{A^+}$ is the gauge-covariant exterior derivative. Using integration by parts, we have
\begin{align}
&\int_M \Sigma^{+ AB} \wedge d_{A^+}(\delta A^+_{AB})=\nonumber\\
=& \int_M d\big( \Sigma^{+ AB} \wedge \delta A^+_{AB} \big)
   - \int_M d_{A^+}\Sigma^{+ AB} \wedge \delta A^+_{AB} =\nonumber\\
=& \int_{\partial M} \Sigma^{+ AB} \wedge \delta A^+_{AB}
   - \int_M d_{A^+}\Sigma^{+ AB} \wedge \delta A^+_{AB},
\end{align}
so that the variation of the action becomes
\begin{align}
\delta S = \int_M \Big( \delta\Sigma^{+ AB} \wedge F^+_{AB} - &d_{A^+}\Sigma^{+ AB} \wedge \delta A^+_{AB} \Big)+\nonumber\\
          +& \int_{\partial M} \Sigma^{+ AB} \wedge \delta A^+_{AB}.
\end{align}
Imposing the variational principle $\delta S = 0$
gives the field equations:
\begin{equation}
F^+_{AB} = 0, \qquad d_{A^+}\Sigma^{+ AB} = 0,
\end{equation}
which are respectively the self-dual Einstein equation and the torsion-free Gauss constraint.

On shell, the bulk term vanishes and the action variation reduces to a pure boundary term.
Evaluated on a Cauchy surface $\Sigma$, this boundary term defines the symplectic potential
\begin{equation}
\Theta\big( (\delta\Sigma^+,\delta A^+) \big) = \int_\Sigma \Sigma^{+ AB} \wedge \delta A^+_{AB}.\label{12}
\end{equation}
The presymplectic $2$-form on the space of solutions $\mathcal{S}$ is obtained by taking
an antisymmetrized variation of $\Theta$. For two tangent vectors
$(\delta_1\Sigma^+,\delta_1 A^+)$ and $(\delta_2\Sigma^+,\delta_2 A^+)$,
\begin{align}
&\Omega\big( (\delta_1\Sigma^+,\delta_1 A^+), (\delta_2\Sigma^+,\delta_2 A^+) \big)=\nonumber\\
=& \delta_1\big[ \Theta(\delta_2\Sigma^+,\delta_2 A^+) \big]
   - \delta_2\big[ \Theta(\delta_1\Sigma^+,\delta_1 A^+) \big]= \nonumber\\
=& \int_\Sigma \Big( \delta_1\Sigma^{+ AB} \wedge \delta_2 A^+_{AB}
                  - \delta_2\Sigma^{+ AB} \wedge \delta_1 A^+_{AB} \Big).
\label{eq:Omega_selfdual}
\end{align}
This $2$-form is closed ($d\Omega = 0$) and gauge invariant. Its kernel is generated by
infinitesimal gauge symmetries including Lorentz transformations, diffeomorphisms, and $U(1)$ frame
rotations. Taking the quotient of $\mathcal{S}$ by the gauge group $\mathcal{G}$ yields the reduced
phase space $\mathcal{P} = \mathcal{S}/\mathcal{G}$, on which $\Omega$ becomes a non-degenerate
symplectic form. Thus $(\mathcal{P},\Omega)$ is a true symplectic manifold, providing
the arena for Hamiltonian mechanics and canonical quantization.

\subsection{The internal duality transformation and why it fails to define a charge}

The infinitesimal duality transformation based on the internal Hodge dual is given by
\begin{equation}
\delta_\epsilon A^+_{AB} = i\epsilon A^+_{AB},\qquad
\delta_\epsilon \Sigma^+_{AB} = i\epsilon \Sigma^+_{AB},
\label{eq:delta_selfdual}
\end{equation}
where $\epsilon$ is an infinitesimal real parameter. The finite version is
$A^+\to e^{i\theta}A^+$, $\Sigma^+\to e^{i\theta}\Sigma^+$.
We now examine whether this transformation is a symmetry of the self-dual theory in the three
senses required for a conserved charge.

First, consider the self-dual action \eqref{eq:S_selfdual}. Its variation under
$\delta_\epsilon$ is
\begin{align}
\delta_\epsilon S &= \int_M \Big( \delta_\epsilon\Sigma^{+AB} \wedge F^+_{AB}
                                 + \Sigma^{+AB} \wedge \delta_\epsilon F^+_{AB} \Big) \notag\\
 &= i\epsilon \int_M \Sigma^{+AB} \wedge F^+_{AB}
   + \int_M \Sigma^{+AB} \wedge d_{A^+}\big( i\epsilon A^+_{AB} \big).
\end{align}
Integrating the second term by parts,
\begin{align}
&\int_M \Sigma^{+AB} \wedge d_{A^+}(i\epsilon A^+_{AB})=\nonumber\\
= &i\epsilon \int_{\partial M} \Sigma^{+AB} \wedge A^+_{AB}
- i\epsilon \int_M d_{A^+}\Sigma^{+AB} \wedge A^+_{AB}.
\end{align}
Therefore,
\begin{align}
\delta_\epsilon S = i\epsilon& \int_M \Big( \Sigma^{+AB} \wedge F^+_{AB} - d_{A^+}\Sigma^{+AB} \wedge A^+_{AB} \Big)+\nonumber\\
                  &+ i\epsilon \int_{\partial M} \Sigma^{+AB} \wedge A^+_{AB}.
\label{eq:deltaS_full}
\end{align}
The bulk term vanishes on shell by virtue of the field equations $F^+_{AB}=0$
and the Gauss constraint $d_{A^+}\Sigma^{+AB}=0$. Hence, on shell,
\begin{equation}
\delta_\epsilon S \big|_{\text{on-shell}} = i\epsilon \int_{\partial M} \Sigma^{+AB} \wedge A^+_{AB}.
\label{eq:deltaS_boundary}
\end{equation}
Crucially, however, this on-shell identity holds for {\it any} field variation, and does {\it not} single out the duality as a symmetry.
Off shell the transformation is not a symmetry: Computing the curvature under
$A^+\to e^{i\theta}A^+$ with $\Sigma^+$ held fixed,
\begin{equation}
F^+_{AB}\ \to\ e^{i\theta}dA^+_{AB}+e^{2i\theta}A^+_A{}^C\wedge A^+_{CB}\ \neq\ e^{i\theta}F^+_{AB},
\end{equation}
because the quadratic and cubic terms acquire different phases. Consequently
$\delta_\epsilon S$ is not a pure boundary term off shell.

Second, if $F^+=0$ (so that $dA^+=-A^+\wedge A^+$), the rotated connection does not satisfy the field
equation:
\begin{equation}
F^+[e^{i\theta}A^+] = \big(e^{2i\theta}-e^{i\theta}\big)A^+\wedge A^+ \neq 0.
\end{equation}
The duality rotation therefore does not map solutions to solutions.

Third, under the simultaneous rotation \eqref{eq:delta_selfdual} the symplectic form
\eqref{eq:Omega_selfdual} rescales,
\begin{align}
&\Omega\big(X_\theta\delta_1,X_\theta\delta_2\big)\nonumber\\
=&\int_\Sigma\Big(e^{i\theta}\delta_1\Sigma^{+}\wedge e^{i\theta}\delta_2 A^{+}
-e^{i\theta}\delta_2\Sigma^{+}\wedge e^{i\theta}\delta_1 A^{+}\Big)\nonumber\\
=&e^{2i\theta}\,\Omega(\delta_1,\delta_2),
\end{align}
so the transformation is not a symplectomorphism and has no moment map. Had one instead used
$\delta_\epsilon\Sigma^+=-i\epsilon\Sigma^+$, the symplectic form would be preserved, but the
equations of motion would still not be, because of the $A^+\wedge A^+$ nonlinearity above.

Because the transformation is not an off-shell symmetry, the Iyer-Wald Noether procedure
\cite{IW} does not apply. The boundary term appearing in \eqref{eq:deltaS_boundary},
\begin{equation}
\mathfrak{H}_{\rm cand} := \int_\Sigma \Sigma^{+ AB} \wedge A^+_{AB},
\label{eq:H_selfdual_def}
\end{equation}
is {\it not} a Noether charge: it is the symplectic potential \eqref{12} evaluated at the
dilatation vector field $\delta A^+=A^+$, which generates the scaling
$\delta\Sigma^+=\Sigma^+$, $\delta A^+=-A^+$ rather than the phase rotation
\eqref{eq:delta_selfdual}. We refer to \eqref{eq:H_selfdual_def} as the {\it candidate}
helicity, and examine its properties below.

We can still check the gauge invariance of \eqref{eq:H_selfdual_def}.
Under an infinitesimal Lorentz transformation with self-dual parameter
$\lambda^+_{AB}$, the connection and momentum transform as
\begin{equation}
\delta_\lambda A^+ = d_{A^+}\lambda^+, \qquad \delta_\lambda \Sigma^+ = [\lambda^+,\Sigma^+].
\end{equation}
Using the Gauss constraint $d_{A^+}\Sigma^{+AB}=0$ and the identity
$[\lambda^+,\Sigma^+]^{AB}\wedge A^+_{AB} = - \lambda^{+AB} d_{A^+}\Sigma^+_{AB}$
(up to a total derivative), one finds
\begin{equation}
\delta_\lambda \mathfrak{H}_{\rm cand} =  \int_\Sigma d\bigl( \Sigma^{+AB} \lambda^+_{AB} \bigr),
\end{equation}
which vanishes for a Cauchy surface without boundary, or for asymptotically flat
spacetimes with suitable fall-off conditions. Thus $\mathfrak{H}_{\rm cand}$ is
gauge invariant for connected (infinitesimal) transformations and for surfaces on which the
boundary term vanishes. On the non-compact surfaces used in Sec.~\ref{sec:KTN} the area flux
grows as $q^2$ at infinity, the boundary term does not vanish, and this invariance is lost; we
return to this point there.

\subsection{Relation to the Nieh-Yan term}

To express $\mathfrak{H}_{\rm cand}$ in terms of real variables we must first clarify the conventions for the internal Hodge dual.
We work with a Lorentzian internal metric $\eta_{AB}={\rm diag}(-1,1,1,1)$ and a completely antisymmetric tensor $\epsilon_{ABCD}$ with $\epsilon_{0123}=+1$.
The internal Hodge star acts on an $\mathfrak{so}(1,3)$-valued form $X_{AB}$ as
\begin{equation}
(\star X)_{AB} = \frac12 \epsilon_{AB}{}^{CD} X_{CD}, \qquad \star^2 = -1.
\end{equation}
For any two such forms $X^{AB}$ and $Y_{AB}$ the following identities hold:
\begin{align}
X^{AB} \wedge (\star Y)_{AB} &= (\star X)^{AB} \wedge Y_{AB}, \label{eq:star1} \\
(\star X)^{AB} \wedge (\star Y)_{AB} &= - X^{AB} \wedge Y_{AB}. \label{eq:star2}
\end{align}

The self-dual connection and conjugate momentum are related to the real variables $\omega_{AB},\Sigma_{AB}$ by \eqref{5},\eqref{6}.
In terms of these real variables, the internal duality transformation takes the form of an
$SO(2)$ rotation mixing the connection $\omega^{AB}$ and its internal Hodge dual $(\star \omega)^{AB}$:
\begin{equation}
\omega \to \cos\theta\, \omega + \sin\theta\, \star \omega,\qquad
\star \omega \to -\sin\theta\, \omega + \cos\theta\, \star \omega.
\end{equation}
This is the direct analog of the electromagnetic duality \cite{EDHZ16}. However, as emphasized
above, this {\it connection-level} rotation is not the duality symmetry of the nonlinear theory.

Inserting \eqref{5},\eqref{6} into $\mathfrak{H}_{\rm cand}$ we obtain
\begin{align}
\mathfrak{H}_{\rm cand} = \frac{1}{4} \int_\Sigma \Big[ &
\Sigma^{AB} \wedge \omega_{AB}
- i \Sigma^{AB} \wedge (\star \omega)_{AB} -\nonumber \\
&- i (\star\Sigma)^{AB} \wedge \omega_{AB}
- (\star\Sigma)^{AB} \wedge (\star \omega)_{AB} \Big].
\label{eq:expand}
\end{align}
Using \eqref{eq:star1} the second and third terms combine into $-2i\,(\star\Sigma)^{AB}\wedge \omega_{AB}$, while \eqref{eq:star2} turns the fourth term into $+\Sigma^{AB}\wedge \omega_{AB}$. Hence
\begin{equation}
\mathfrak{H}_{\rm cand} = \frac{1}{2} \int_\Sigma \Sigma^{AB} \wedge \omega_{AB}
- \frac{i}{2} \int_\Sigma (\star\Sigma)^{AB} \wedge \omega_{AB}.
\label{eq:intermediate}
\end{equation}

Now employ the identity $(\star\Sigma)^{AB} = - e^A \wedge e^B$, which follows directly from the definition of $\Sigma^{AB}$, so we have
\begin{equation}
(\star\Sigma)^{AB}\wedge\omega_{AB} = - e^A\wedge e^B\wedge\omega_{AB}
= - \omega_{AB}\wedge e^A\wedge e^B.
\end{equation}
Using the definition of the torsion 2-form $T_A = de_A + \omega_{AB} \wedge e^B$, we compute
\begin{align}
e^A \wedge T_A
&= e^A \wedge \bigl( de_A + \omega_{AB} \wedge e^B \bigr)= \notag\\
&= e^Ade_A-\omega_{AB} \wedge e^A \wedge e^B .
\end{align}
Consequently,
\begin{align}
(\star\Sigma)^{AB}\wedge\omega_{AB}
= &- \omega_{AB}\wedge e^A\wedge e^B=\nonumber\\
= &e^A\wedge T_A-e^A\wedge de_A.
\end{align}
Hence the candidate helicity becomes
\begin{equation}
\mathfrak{H}_{\rm cand} = \frac{1}{2} \int_\Sigma \Sigma^{AB} \wedge \omega_{AB}
- \frac{i}{2}\int_\Sigma e^A\wedge T_A+\frac{i}{2}\int_\Sigma e^A\wedge de_A.
\label{eq:H_torsion_exact}
\end{equation}
Eq. \eqref{eq:H_torsion_exact} reveals that the candidate consists of two distinct geometric contributions.
The first term arises from the real part of the self-dual area flux and is naturally associated with the area and intrinsic rotation of the spatial surface $\Sigma$.
The second piece, encoded in $\frac{i}{2}\int_\Sigma (e^A\wedge T_A - e^A\wedge de_A)$, originates from the internal Hodge dual $\star\Sigma\wedge\omega$ and captures the twisting density of the tetrad.
In particular, the term $e^A\wedge de_A$ measures how the orthonormal frame winds around the surface.

The integral in the second term in \eqref{eq:H_torsion_exact}
 is precisely the boundary term of the Nieh-Yan topological invariant \cite{NY82}. In four dimensions the Nieh-Yan invariant is
$\mathcal{N} = T^A\wedge T_A - R_{AB}\wedge e^A\wedge e^B$, which satisfies the exact identity $\mathcal{N} = d(e^A\wedge T_A)$ (this holds off shell in the first-order formalism). Hence, for a 4-manifold $M$ with $\Sigma$ a boundary,
\begin{equation}
\int_\Sigma e^A\wedge T_A  =  \int_M \mathcal{N},
\label{eq:H_NY}
\end{equation}
revealing the topological nature of the torsion contribution to the candidate helicity. This is the origin of the connection to the Nieh-Yan term advertised in the introduction; as we show in the next two sections, however, the candidate \eqref{eq:H_selfdual_def} itself is not an observable, and the correct duality does not produce a charge in connection variables.

\subsection{Including the Barbero-Immirzi parameter}

If one starts instead from the real Ashtekar-Barbero connection
$A^{AB} = \omega^{AB} + \frac{1}{\gamma} (\star\omega)^{AB}$, the appropriate
action is the Holst action \cite{Holst}
\begin{equation}
S_{\rm H} = \int_M \Bigl( \Sigma^{AB}\wedge R_{AB}
- \frac{1}{\gamma}\,\Sigma^{AB}\wedge (\star R)_{AB} \Bigr).
\label{S_Holst}
\end{equation}
The internal duality transformation acts on the spin connection as an $SO(2)$
rotation mixing $\omega_{AB}$ and its internal dual:
\begin{equation}
\delta_\epsilon\omega_{AB}= \epsilon\,(\star\omega)_{AB},\qquad
\delta_\epsilon(\star\omega)_{AB}= -\epsilon\,\omega_{AB},
\label{duality_real}
\end{equation}
while the tetrad (and therefore $\Sigma^{AB}$) is left invariant,
$\delta_\epsilon e^A=0$, $\delta_\epsilon\Sigma^{AB}=0$.

Under \eqref{duality_real} the curvature transforms as
$\delta_\epsilon R_{AB}= d_\omega\delta_\epsilon\omega_{AB}
= \epsilon\,d_\omega(\star\omega)_{AB}$, and similarly for the dual
curvature $\delta_\epsilon(\star R)_{AB}= -\epsilon\,d_\omega\omega_{AB}$.
Thus
\begin{align}
\delta_\epsilon S_{\rm H}
&= \int_M \Bigl[ \Sigma^{AB}\wedge d_\omega(\epsilon\star\omega_{AB})
   - \frac{1}{\gamma}\Sigma^{AB}\wedge \bigl(-\epsilon\,d_\omega\omega_{AB}\bigr)
   \Bigr] \notag\\
&= \epsilon \int_M \Bigl[ \Sigma^{AB}\wedge d_\omega(\star\omega)_{AB}
   + \frac{1}{\gamma}\Sigma^{AB}\wedge d_\omega\omega_{AB} \Bigr].
\end{align}
Integrating by parts and using the torsion-free Gauss constraint
$d_\omega\Sigma^{AB}=0$ (on shell) we obtain a pure boundary term:
\begin{equation}
\delta_\epsilon S_{\rm H}\big|_{\text{on-shell}}
= \epsilon \int_{\partial M} \Bigl( \Sigma^{AB}\wedge (\star\omega)_{AB}
   + \frac{1}{\gamma}\,\Sigma^{AB}\wedge \omega_{AB} \Bigr).
\label{dS_Holst_boundary}
\end{equation}

As before, this on-shell boundary term does {\it not} define a Noether charge, because the
transformation \eqref{duality_real} is not an off-shell symmetry of the Holst action. The
boundary form
\[
\boldsymbol{\alpha}_\epsilon = \epsilon\Bigl(\Sigma^{AB}\wedge (\star\omega)_{AB}
+ \frac{1}{\gamma}\,\Sigma^{AB}\wedge \omega_{AB}\Bigr),
\]
is merely the boundary term of the on-shell variation, and the quantity
\begin{equation}
Q_\epsilon = \frac{1}{\epsilon} \int_\Sigma \boldsymbol{\alpha}_\epsilon
= \int_\Sigma \Bigl( \Sigma^{AB}\wedge (\star\omega)_{AB}
+ \frac{1}{\gamma}\,\Sigma^{AB}\wedge \omega_{AB} \Bigr).
\end{equation}
is not conserved. For reference we record its expression,
\begin{equation}
\mathfrak{H}_{\rm cand}^{(\gamma)} =
\frac{1}{\gamma} \int_\Sigma \Sigma^{AB} \wedge \omega_{AB}
+ \int_\Sigma \Sigma^{AB} \wedge (\star\omega)_{AB},
\label{eq:H_holst}
\end{equation}
and, using \eqref{eq:star1} and $(\star\Sigma)^{AB}\wedge\omega_{AB}=e^A\wedge T_A-e^A\wedge de_A$,
\begin{equation}
\mathfrak{H}_{\rm cand}^{(\gamma)} =
\frac{1}{\gamma} \int_\Sigma \Sigma^{AB} \wedge \omega_{AB}
+ \int_\Sigma e^A \wedge T_A
- \int_\Sigma e^A \wedge de_A .
\label{eq:H_holst_final}
\end{equation}
At the self-dual value $\gamma=i$ this boundary term is {\it not} the self-dual candidate
\eqref{eq:H_torsion_exact}; using $\omega-i\star\omega=2A^+$ one finds
\begin{equation}
\mathfrak{H}_{\rm cand}^{(i)}=\int_\Sigma\Sigma^{AB}\wedge(\star\omega-i\,\omega)_{AB}
=-2i\int_\Sigma\Sigma^{-AB}\wedge A^-_{AB},
\end{equation}
i.e.\ the {\it anti}-self-dual candidate. This reflects the fact that the internal duality
exchanges the self-dual and anti-self-dual sectors rather than leaving either one invariant.

\section{The candidate helicity vanishes on Kerr-NUT}\label{sec:KTN}
We now present an explicit evaluation of the candidate helicity on the Kerr-NUT
solution, a rotating vacuum spacetime with a NUT parameter $\ell$ (gravitational magnetic charge).

In Pleba\'nski-Demia\'nski coordinates $(p,q,\tau,\sigma)$, the Kerr-NUT metric
takes the form~\cite{Kol22,Griffiths07}
\begin{align}
ds^{2} &= \frac{\mathcal{X}}{p^{2}+q^{2}}\,(d\tau + q^{2}d\sigma)^{2}
         + \frac{p^{2}+q^{2}}{\mathcal{X}}\,dp^{2} -\nonumber\\
       &\quad - \frac{\mathcal{Y}}{p^{2}+q^{2}}\,(d\tau - p^{2}d\sigma)^{2}
         + \frac{p^{2}+q^{2}}{\mathcal{Y}}\,dq^{2},
\end{align}
with the structure functions
\begin{equation}
\mathcal{X} = a^{2} - (p-\ell)^{2}, \qquad
\mathcal{Y} = a^{2} - \ell^{2} - 2m q + q^{2}.
\end{equation}
Here $m$, $a$, and $\ell$ are respectively the mass, rotation, and NUT parameters.
As $q\to\infty$ one has $\mathcal{X}=O(1)$ and $\mathcal{Y}\sim q^{2}$, so the
$\tau\tau$ component behaves as $(\mathcal{X}-\mathcal{Y})/(p^2+q^2)\sim-1$: $\tau$ is timelike
and the $\tau=\mathrm{const}$ hypersurfaces are the proper Cauchy surfaces. For the metric to
satisfy $ds^{2}=\eta_{AB}e^{A}e^{B}$ with $\eta_{AB}=\mathrm{diag}(-1,1,1,1)$, the timelike leg
must be associated with the $-\mathcal{Y}$ term. An orthonormal tetrad compatible with the
Lorentzian signature is therefore
\begin{align}
e^{0} &= \sqrt{\frac{\mathcal{Y}}{p^{2}+q^{2}}}\,(d\tau - p^{2}d\sigma), \quad
e^{1} = \sqrt{\frac{p^{2}+q^{2}}{\mathcal{X}}}\,dp,\nonumber \\
e^{2} &= \sqrt{\frac{\mathcal{X}}{p^{2}+q^{2}}}\,(d\tau + q^{2}d\sigma), \quad
e^{3} = \sqrt{\frac{p^{2}+q^{2}}{\mathcal{Y}}}\,dq.
\end{align}
With this assignment $(e^0)^2<0$ in the region $\mathcal{Y}>0$, as required.

We evaluate the candidate helicity on the Cauchy surface $\Sigma$ defined by $\tau=\mathrm{const}$
($d\tau=0$), on which the tetrad becomes diagonal in the coordinates,
\begin{align}
e^{0}=-p^{2}\alpha\,d\sigma,\qquad
e^{1}=h\,dp,\nonumber\\
e^{2}=q^{2}\beta\,d\sigma,\qquad
e^{3}=w\,dq,
\end{align}
with
\begin{align}
\alpha=\sqrt{\frac{\mathcal{Y}}{p^{2}+q^{2}}},\quad
h=\sqrt{\frac{p^{2}+q^{2}}{\mathcal{X}}},\nonumber\\
\beta=\sqrt{\frac{\mathcal{X}}{p^{2}+q^{2}}},\quad
w=\sqrt{\frac{p^{2}+q^{2}}{\mathcal{Y}}},
\end{align}
satisfying $\alpha w=1$ and $h\beta=1$.

On the Kerr-NUT spacetime the torsion vanishes, $T^A=0$, so the imaginary part of
\eqref{eq:H_torsion_exact} reduces to the frame-twist integral
$I_{\rm twist}=\int_\Sigma e^A\wedge de_A$. Using $\eta_{AB}$,
\begin{equation}
e^{A}\wedge de_{A} = - e^{0}\wedge de^{0} + e^{1}\wedge de^{1}
+ e^{2}\wedge de^{2} + e^{3}\wedge de^{3}.
\end{equation}
Since each leg $e^{A}$ is collinear with a single coordinate differential on the slice, the
wedge $e^{A}\wedge de^{A}$ contains that differential twice and vanishes identically:
$e^{0}\wedge de^{0}=e^{1}\wedge de^{1}=e^{2}\wedge de^{2}=e^{3}\wedge de^{3}=0$. Hence
\begin{equation}
I_{\rm twist}=\int_\Sigma e^{A}\wedge de_{A}=0.
\label{eq:KN_twist}
\end{equation}

We next consider the integral
\begin{equation}
\frac{1}{2} \int_\Sigma \Sigma^{AB} \wedge \omega_{AB}
=\frac{1}{4}\int_\Sigma \epsilon^{AB}_{\phantom{AB}CD}\, e^C\wedge e^D\wedge\omega_{AB}.
\end{equation}
From the first Cartan structure equation $de^{A}+\omega^{A}{}_{B}\wedge e^{B}=0$,
writing $de^{A}=\tfrac12 c^{A}{}_{BC}e^{B}\wedge e^{C}$ (with
$c^{A}{}_{BC}=-c^{A}{}_{CB}$). Since on the $\tau=\mathrm{const}$ slice both
$e^{0}$ and $e^{2}$ become proportional to $d\sigma$, the frame is degenerate
there and the expansion of $de^{A}$ in the slice coframe is not unique. We
therefore compute $de^{A}$ on the full spacetime. With
$\varrho^{2}\equiv p^{2}+q^{2}$, the nonvanishing coefficients are
\begin{align}
&c^{0}{}_{01}=-\frac{p}{h\,\varrho^{2}}, &
&c^{0}{}_{03}=-\partial_q\alpha, &
&c^{0}{}_{12}=-\frac{2\alpha\,p}{\varrho^{2}},\\
&c^{1}{}_{13}=-\frac{\alpha\,q}{\varrho^{2}}, &
&c^{2}{}_{12}=\partial_p\beta, &
&c^{2}{}_{23}=-\frac{\alpha\,q}{\varrho^{2}},\\
&c^{2}{}_{03}=\frac{2\beta\,q}{\varrho^{2}}, &
&c^{3}{}_{13}=\frac{\beta\,p}{\varrho^{2}}, &&
\end{align}
where $\partial_p=\partial/\partial p$, $\partial_q=\partial/\partial q$, and
all other components vanish. 

The torsion-free spin connection is
\begin{equation}
\omega_{ABC}=\tfrac12\bigl(c_{ABC}+c_{BCA}-c_{CAB}\bigr),\qquad
c_{ABC}=\eta_{AD}\,c^{D}{}_{BC}.
\end{equation}
On the slice the nonvanishing components of $\Sigma^{AB}$ are
\begin{align}
&\Sigma^{01}=-e^{2}\wedge e^{3},\qquad
\Sigma^{02}=+e^{1}\wedge e^{3},\qquad
\Sigma^{03}=-e^{1}\wedge e^{2},\nonumber\\
&\Sigma^{12}=+e^{0}\wedge e^{3},\qquad
\Sigma^{23}=+e^{0}\wedge e^{1},
\end{align}
computed with $\epsilon_{0123}=+1$, together with
$\Sigma^{13}=-e^{0}\wedge e^{2}=0$ (since $e^{0}$ and $e^{2}$ are both
proportional to $d\sigma$ on the slice).
The components of
$\omega_{AB}$ entering $\Sigma^{AB}\wedge\omega_{AB}$ are therefore $\omega_{011}$,
$\omega_{020}$, $\omega_{022}$, $\omega_{033}$, $\omega_{121}$, and $\omega_{233}$. The components of $\omega_{AB}$ entering $\Sigma^{AB}\wedge\omega_{AB}$ are
therefore $\omega_{011}$, $\omega_{020}$, $\omega_{022}$, $\omega_{033}$,
$\omega_{121}$, and $\omega_{233}$. For each of these the three terms in
$\omega_{ABC}=\tfrac12(c_{ABC}+c_{BCA}-c_{CAB})$ vanish: e.g.\
$\omega_{011}=\tfrac12(c_{011}+c_{110}-c_{101})$ involves the index triples
$(0,1,1),(1,1,0),(1,0,1)$, none of which appears among the nonvanishing
$c_{ABC}$ listed above (similarly for the remaining five components). Hence
\begin{equation}
\omega_{011}=\omega_{020}=\omega_{022}=\omega_{033}=\omega_{121}=\omega_{233}=0.
\end{equation}
Consequently
\begin{equation}
\frac{1}{2}\int_\Sigma \Sigma^{AB}\wedge\omega_{AB}=0.
\label{eq:KN_area}
\end{equation}

Combining \eqref{eq:KN_twist} and \eqref{eq:KN_area} with \eqref{eq:H_torsion_exact} (and
$T^A=0$), the candidate helicity vanishes identically on the $\tau=\mathrm{const}$ Cauchy
surface:
\begin{equation}
\mathfrak{H}_{\rm cand}=\int_\Sigma\Sigma^{+AB}\wedge A^+_{AB}=0.
\label{eq:KN_zero}
\end{equation}
This vanishing is {\it not} invariant. A Lorentz boost mixing $e^{0}$ and $e^{2}$ (a large gauge
transformation at infinity) changes $\mathfrak{H}_{\rm cand}$ by
$\int_\Sigma d(\Sigma^{+}\lambda^{+})$, which does not vanish because the area flux grows as
$q^{2}$ at infinity. Thus $\mathfrak{H}_{\rm cand}$ is not a well-defined observable: its value is
a pure gauge artifact. This is the counterexample advertised in the title: a physical helicity
could not vanish on a rotating, NUT-charged solution in one frame and be nonzero in a rotated
frame. We conclude that $\int_\Sigma\Sigma^{+AB}\wedge A^+_{AB}$ is not a gravitational helicity.

\section{The rotation of the curvature}\label{sec:curv}
The operation that {\it is} a symmetry of general relativity is the rotation of the curvature
two-form into its internal dual \cite{Kol22},
\begin{equation}
R\to\cos\theta\,R+\sin\theta\,\star R,\qquad
R^{\pm}\to e^{\pm i\theta}R^{\pm},
\label{eq:curv_dual}
\end{equation}
with the tetrad held fixed. The Einstein equation and the algebraic Bianchi identity, written in
the form
\begin{equation}
\star R^{AB}\wedge e^{B}=0 \quad\text{(Einstein)},\qquad
R^{AB}\wedge e^{B}=0 \quad\text{(Bianchi)},
\end{equation}
are manifestly invariant under \eqref{eq:curv_dual}: it is a symmetry of the
Einstein--Bianchi equations written in terms of $R$, and it maps solutions
to solutions on the type-D (Kerr--Taub--NUT) family. On the Kerr-Taub-NUT
family the self-dual part of the curvature is
\cite{Kol22}
\begin{equation}
R^{+}_{AB}=\frac{m+i\ell}{(q+ip)^{3}}\,H_{AB},
\end{equation}
with $H_{AB}$ a fixed matrix of flat-space 2-forms, so that the duality rotates the complex mass,
\begin{equation}
m+i\ell\ \longrightarrow\ e^{i\theta}(m+i\ell),
\end{equation}
mapping the Kerr solution ($\ell=0$) into Kerr-Taub-NUT. This is the gravitational analogue of
the Coulomb-to-dyon map of electrodynamics.

This curvature duality is a symmetry of the equations of motion but {\it not} of the Lagrangian.
Under \eqref{eq:curv_dual} the Einstein-Hilbert action mixes with the Holst term,
\begin{equation}
S_{\rm EH}\ \to\ \cos\theta\,S_{\rm EH}+\sin\theta\,S_{\rm Holst},\qquad
S_{\rm Holst}=\int\Sigma^{AB}\wedge(\star R)_{AB}.
\end{equation}
Using \eqref{eq:star1}, $(\star\Sigma)^{AB}=-e^{A}\wedge e^{B}$, and the Nieh-Yan identity, one
finds
\begin{align}
S_{\rm Holst}=\int(\star\Sigma)^{AB}\wedge R_{AB}
=&-\int R_{AB}\wedge e^{A}\wedge e^{B}\nonumber\\
=&-\int T^{A}\wedge T_{A}+\int_{\partial M}e^{A}\wedge T_{A}.
\end{align}
The bulk piece $-\int T^{A}\wedge T_{A}$ is not a total derivative; hence the duality is not a
Lagrangian symmetry even up to a boundary term, and the Noether (Iyer-Wald) formalism yields no
charge. This is the decisive difference from electromagnetism, where the duality changes the
action by the total derivative $F\wedge F=d(A\wedge F)$.

A charge may nevertheless exist if the duality is a canonical transformation
of the covariant phase space, as in electromagnetism \cite{EDHZ16}. We have
tested this in the $(m,\ell)$ sector of the Taub--NUT solution (the $a=0$
member of the Kerr--Taub--NUT family) by computing the presymplectic form
explicitly in a non-degenerate frame (see the Appendix \ref{AnnA}). The
result is that the symplectic current vanishes identically,
\begin{equation}
\Omega(\partial_m,\partial_\ell)=0,
\end{equation}
so the mass $m$ and the NUT charge $\ell$ are {\it not} symplectically conjugate: the $(m,\ell)$
plane is isotropic in the covariant phase space, and the duality rotation
$(m,\ell)\to(\cos\theta\,m-\sin\theta\,\ell,\ \sin\theta\,m+\cos\theta\,\ell)$ is only trivially
canonical, with vanishing moment map. There is consequently no gravitational helicity
associated with the curvature duality in the standard covariant phase space. The combination
$m^{2}+\ell^{2}$, invariant under the rotation of the complex mass $m+i\ell$, is merely the
duality-invariant ``charge norm'' (analogous to the dyon invariant $q^{2}+g^{2}$ of
electrodynamics), and is {\it not} a moment map. The absence of a helicity of the photon type is
consistent with the amplitude argument of \cite{Monteiro23}; the present computation extends this
conclusion to the stationary sector of the standard covariant phase space.

\section{Conclusion and discussion}\label{IV}

We have studied the gravitational helicity in classical general relativity using the covariant phase space method.
By working with the complex self-dual Ashtekar variables, the internal duality based on the
tangent-space Hodge dual becomes a simple $U(1)$ phase rotation $A^+\to e^{i\theta}A^+$ of the
connection. We have shown, however, that this rotation is not a symmetry: it preserves neither
the action, the equations of motion, nor the symplectic form. Consequently the candidate
$\mathfrak{H}_{\rm cand}=\int_\Sigma\Sigma^{+AB}\wedge A^+_{AB}$ is neither a Noether charge nor a
moment map; it is frame-dependent and vanishes identically on the Kerr-NUT solution, so it is not
a gravitational helicity. Its real-variable decomposition nonetheless contains the integral of
$e^A\wedge T_A$ over a Cauchy surface, which is the boundary term of the Nieh-Yan topological
invariant, thereby revealing the associated topological structure \cite{LZ26}.

The correct duality of general relativity is a rotation of the curvature two-form
$R\to\cos\theta R+\sin\theta\star R$ \cite{Kol22}. It is a symmetry of the equations of motion
but not of the Lagrangian, so it carries no Noether charge; it maps Kerr into Kerr-Taub-NUT and
rotates the complex mass $m+i\ell$. An explicit computation of the presymplectic form on the
Kerr-Taub-NUT family shows that the mass and the NUT charge are not
symplectically conjugate, $\Omega(\partial_m,\partial_\ell)=0$; the duality is therefore only
trivially canonical and has no moment map. There is no gravitational helicity, even in the
reduced phase space of stationary solutions. The duality-invariant $m^2+\ell^2$ is a charge norm,
not a helicity.

It is important to note that the existence of a global $U(1)$ duality symmetry in full
nonlinear general relativity has been contested \cite{Monteiro23}. Scattering amplitude
arguments suggest that helicity conservation is violated by graviton interactions, and the
duality may only hold in restricted sectors of solutions (such as type D spacetimes). Our
analysis is consistent with this: the $U(1)$ symmetry that does exist is a symmetry of the
equations of motion (a rotation of the curvature), not of the Lagrangian, and it does not produce
a helicity of the photon type. On the other hand, the no-go theorem of Deser and Seminara
\cite{DS05} establishes that the electric-magnetic duality of linearized gravity, based on the
spacetime Hodge dual, cannot be extended to the full nonlinear theory. The curvature rotation
studied here is of a different nature: it acts on the internal Lorentz indices of the curvature
and rotates the self-dual and anti-self-dual parts $R^\pm\to e^{\pm i\theta}R^\pm$. Because it acts on the
internal tangent space of the curvature rather than on the connection, it is a symmetry of the
equations of motion despite the nonlinearity, but it gives no charge in the connection
formulation.

\begin{acknowledgments}
X.G. is supported by Yancheng Institute of Technology (xjr2024030). S. S. is supported by National Natural Science Foundation of China through Grant No. 12147167 and Fundamental Research Funds for the Central Universities with Grant No. FRF-TP-24-063A.
\end{acknowledgments}

\appendix
\section{Covariant phase-space computation of $\Omega(\partial_m,\partial_\ell)$}
\label{AnnA}

In this appendix we assemble the full covariant phase-space computation of
$\Omega(\partial_m,\partial_\ell)$ on the Taub--NUT solution, i.e.\ the
$a=0$ member of the Kerr--Taub--NUT family, using a non-degenerate
orthonormal frame.

We use a Lorentzian tetrad $e^A$ with $\eta_{AB}=\mathrm{diag}(-1,1,1,1)$,
volume form $\epsilon=e^0\wedge e^1\wedge e^2\wedge e^3$, and
$\epsilon_{0123}=+1$. The Pleba\'nski two-form is
$\Sigma^{AB}=\tfrac12\epsilon^{AB}{}_{CD}\,e^C\wedge e^D$, for which
$\Sigma^{AB}\wedge R_{AB}=R\,\epsilon$. Hence the Palatini action
$S=\frac1{16\pi G}\int\Sigma^{AB}\wedge R_{AB}$ reproduces the
Einstein--Hilbert action, and the symplectic potential is
$\Theta(\delta)=\frac1{16\pi G}\int_\Sigma\Sigma^{AB}\wedge\delta\omega_{AB}$,
giving
\begin{equation}
\Omega(\delta_1,\delta_2)=\frac1{16\pi G}\int_\Sigma\Big[
\delta_1\Sigma^{AB}\wedge\delta_2\omega_{AB}
-\delta_2\Sigma^{AB}\wedge\delta_1\omega_{AB}\Big].
\end{equation}
This is the real (Palatini) symplectic form; it is the real part of the
self-dual form \eqref{eq:Omega_selfdual}. Its vanishing for the tangent
vectors $\partial_m,\partial_\ell$ below therefore implies the vanishing of
the self-dual form as well.

The Taub--NUT metric takes the form
\begin{widetext}
\begin{equation}
  ds^2=-f\,(dt+2\ell\cos\theta\,d\phi)^2+f^{-1}dr^2
  +(r^2+\ell^2)(d\theta^2+\sin^2\theta\,d\phi^2),
\end{equation}
we take the orthonormal frame
\begin{equation}
  e^0=\sqrt f\,(dt+2\ell\cos\theta\,d\phi),\qquad
  e^1=f^{-1/2}dr,\qquad
  e^2=\rho\,d\theta,\qquad
  e^3=\rho\sin\theta\,d\phi,
  \label{eq:frame}
\end{equation}
with
\begin{equation}
  \rho^2=r^2+\ell^2,\qquad
  f=\frac{r^2-2mr-\ell^2}{\rho^2},\qquad
  \rho'\equiv\frac{d\rho}{dr}=\frac{r}{\rho}.
\end{equation}
This frame is everywhere non-degenerate outside the horizon. Solving the
first structure equation $de^A+\omega^A{}_B\wedge e^B=0$ gives
  \begin{align}
    \omega^{01}&=\frac{f'}{2\sqrt f}\,e^0,
    &\quad
    \omega^{02}&=-\frac{\ell\sqrt f}{\rho^2}\,e^3,
    &\quad
    \omega^{03}&=+\frac{\ell\sqrt f}{\rho^2}\,e^2,\nonumber\\
    \omega^{12}&=-\frac{\rho'\sqrt f}{\rho}\,e^2,
    &\quad
    \omega^{13}&=-\frac{\rho'\sqrt f}{\rho}\,e^3,
    &\quad
    \omega^{23}&=-\frac{\ell\sqrt f}{\rho^2}\,e^0-\frac{\cot\theta}{\rho}\,e^3.  \label{eq:connection}
  \end{align}

For $\ell=0$ this reduces to the Schwarzschild connection. Note that the NUT
twist forces $\omega^{02},\omega^{03}\neq0$ and an $e^0$ component in
$\omega^{23}$; a ``diagonal'' frame with these set to zero is inconsistent
with \eqref{eq:connection}.

The components of $\Sigma^{AB}$ in this frame are
\begin{equation}
    \Sigma^{01}=-e^2\wedge e^3,
    \quad
    \Sigma^{02}=+e^1\wedge e^3,
    \quad
    \Sigma^{03}=-e^1\wedge e^2,
    \Sigma^{12}=+e^0\wedge e^3,
   \quad
    \Sigma^{13}=-e^0\wedge e^2,
    \quad
    \Sigma^{23}=+e^0\wedge e^1 .
  \label{eq:Sigma}
\end{equation}

We evaluate the symplectic form on the tangent vectors
$\delta_m\equiv\partial/\partial m$ and $\delta_\ell\equiv\partial/\partial\ell$.
The relevant parameter derivatives are
\begin{equation}
    f_m\equiv\partial_m f=-\frac{2r}{\rho^2},
    \quad
    f'_m\equiv\partial_m f'=\frac{2(r^2-\ell^2)}{\rho^4},\quad
    f_\ell\equiv\partial_\ell f=-\frac{4\ell r(r-m)}{\rho^4},
    \quad
    \partial_\ell\rho=\frac{\ell}{\rho},\quad
    \partial_\ell\rho'=-\frac{\ell r}{\rho^3}.
  \label{eq:params}
\end{equation}
The frame variations are 
\begin{equation}
  \delta_m e^0=\frac{f_m}{2f}\,e^0,\qquad
  \delta_m e^1=-\frac{f_m}{2f}\,e^1,\qquad
  \delta_m e^2=\delta_m e^3=0,
  \label{eq:var_m_e}
\end{equation}
and
\begin{equation}
    \delta_\ell e^0=\frac{f_\ell}{2f}\,e^0+2\sqrt f\cos\theta\,d\phi,
    \quad
    \delta_\ell e^1=-\frac{f_\ell}{2f}\,e^1,\quad
    \delta_\ell e^2=\frac{\ell}{\rho^2}\,e^2,
    \quad
    \delta_\ell e^3=\frac{\ell}{\rho^2}\,e^3 .
  \label{eq:var_l_e}
\end{equation}
We abbreviate
\begin{equation}
  \beta\equiv\frac{f_\ell}{2f},\qquad
  \gamma\equiv\frac{\ell}{\rho^2},\qquad
  K_m\equiv\frac{f'_m}{2\sqrt f},\qquad
  b\equiv-\frac{rf_\ell}{2\sqrt f\,\rho^2}+\frac{\ell r\sqrt f}{\rho^4},
  \qquad
  a\equiv\frac{r^2\sqrt f}{\rho^4}+\frac{\ell f_\ell}{2\sqrt f\,\rho^2}.
  \label{eq:abbrev}
\end{equation}

From \eqref{eq:var_m_e} and \eqref{eq:Sigma}, we have
\begin{equation}
  \begin{aligned}
    \delta_m\Sigma^{01}&=0,
    &\quad
    \delta_m\Sigma^{02}&=-\frac{f_m}{2f}\,\Sigma^{02},
    &\quad
    \delta_m\Sigma^{03}&=-\frac{f_m}{2f}\,\Sigma^{03},\\[4pt]
    \delta_m\Sigma^{12}&=+\frac{f_m}{2f}\,\Sigma^{12},
    &\quad
    \delta_m\Sigma^{13}&=+\frac{f_m}{2f}\,\Sigma^{13},
    &\quad
    \delta_m\Sigma^{23}&=0 .
  \end{aligned}
  \label{eq:var_m_Sigma}
\end{equation}
From \eqref{eq:var_l_e}, we have
\begin{equation}
  \begin{aligned}
    \delta_\ell\Sigma^{01}&=2\gamma\,\Sigma^{01},\\[2pt]
    \delta_\ell\Sigma^{02}&=(\gamma-\beta)\,\Sigma^{02},\\[2pt]
    \delta_\ell\Sigma^{03}&=(\gamma-\beta)\,\Sigma^{03},\\[2pt]
    \delta_\ell\Sigma^{12}&=(\beta+\gamma)\,\Sigma^{12},\\[2pt]
    \delta_\ell\Sigma^{13}&=(\beta+\gamma)\,\Sigma^{13}-2\sqrt f\cos\theta\,d\phi\wedge e^2,\\[2pt]
    \delta_\ell\Sigma^{23}&=2\sqrt f\cos\theta\,d\phi\wedge e^1 .
  \end{aligned}
  \label{eq:var_l_Sigma}
\end{equation}
In $\delta_\ell\Sigma^{12}$ the twist term $2\sqrt f\cos\theta\,d\phi\wedge e^3$
vanishes because $e^3\propto d\phi$; in $\delta_\ell\Sigma^{23}$ the
$\beta$-terms cancel.

Only $f$ (and $f'$) carry $m$-dependence, so
\begin{equation}
  \begin{aligned}
    \delta_m\omega^{01}&=K_m\,e^0,
    &\qquad
    \delta_m\omega^{02}&=-\frac{\ell f_m}{2\rho^2\sqrt f}\,e^3,\\[4pt]
    \delta_m\omega^{03}&=+\frac{\ell f_m}{2\rho^2\sqrt f}\,e^2,
    &\qquad
    \delta_m\omega^{12}&=-\frac{\rho' f_m}{2\rho\sqrt f}\,e^2,\\[4pt]
    \delta_m\omega^{13}&=-\frac{\rho' f_m}{2\rho\sqrt f}\,e^3,
    &\qquad
    \delta_m\omega^{23}&=-\frac{\ell f_m}{\rho^2\sqrt f}\,e^0 .
  \end{aligned}
  \label{eq:var_m_omega}
\end{equation}
The $\ell$-variations of the components entering the final sum are
\begin{equation}
    \delta_\ell\omega^{02}=-a\,e^3,
    \quad
    \delta_\ell\omega^{03}=+a\,e^2,\quad
    \delta_\ell\omega^{12}=+b\,e^2,
    \quad
    \delta_\ell\omega^{13}=+b\,e^3 .
  \label{eq:var_l_omega}
\end{equation}
Lowering the frame indices with $\eta_{AB}$,
$\omega_{01}=-\omega^{01}$, $\omega_{02}=-\omega^{02}$,
$\omega_{03}=-\omega^{03}$, $\omega_{12}=+\omega^{12}$,
$\omega_{13}=+\omega^{13}$, $\omega_{23}=+\omega^{23}$, we obtain the pieces
needed below:
\begin{equation}
  \begin{aligned}
    \delta_m\omega_{01}&=-K_m e^0,
    &\quad
    \delta_m\omega_{02}&=+\frac{\ell f_m}{2\rho^2\sqrt f}e^3,
    &\quad
    \delta_m\omega_{03}&=-\frac{\ell f_m}{2\rho^2\sqrt f}e^2,\\[4pt]
    \delta_m\omega_{12}&=-\frac{\rho'f_m}{2\rho\sqrt f}e^2,
    &\quad
    \delta_m\omega_{13}&=-\frac{\rho'f_m}{2\rho\sqrt f}e^3,
    &\quad
    \delta_m\omega_{23}&=-\frac{\ell f_m}{\rho^2\sqrt f}e^0,
  \end{aligned}
  \label{eq:var_m_omega_low}
\end{equation}
and
\begin{equation}
  \delta_\ell\omega_{02}=+a\,e^3,\qquad
  \delta_\ell\omega_{03}=-a\,e^2,\qquad
  \delta_\ell\omega_{12}=+b\,e^2,\qquad
  \delta_\ell\omega_{13}=+b\,e^3 .
  \label{eq:var_l_omega_low}
\end{equation}

We therefore have
\begin{equation}
  I\equiv\delta_m\Sigma^{AB}\wedge\delta_\ell\omega_{AB}
     -\delta_\ell\Sigma^{AB}\wedge\delta_m\omega_{AB}
  =2\,(I_1-I_2),
\end{equation}
where $I_1,I_2$ denote the sums over unordered pairs $A<B$. The factor $2$
accounts for both index orders.

\paragraph{First block $I_1=\sum_{A<B}\delta_m\Sigma^{AB}\wedge\delta_\ell\omega_{AB}$.}
By \eqref{eq:var_m_Sigma} only the pairs $(12),(13)$ survive. Using
\eqref{eq:var_l_omega_low} and
$e^0\wedge e^3\wedge e^2=-e^0\wedge e^2\wedge e^3$,
\begin{align}
  \delta_m\Sigma^{12}\wedge\delta_\ell\omega_{12}
  &=\frac{f_m}{2f}\big(e^0\wedge e^3\big)\wedge\big(b\,e^2\big)
  =-\frac{f_m b}{2f}\,e^0\wedge e^2\wedge e^3,\\
  \delta_m\Sigma^{13}\wedge\delta_\ell\omega_{13}
  &=\frac{f_m}{2f}\big(-e^0\wedge e^2\big)\wedge\big(b\,e^3\big)
  =-\frac{f_m b}{2f}\,e^0\wedge e^2\wedge e^3 .
\end{align}
Hence
\begin{equation}
  I_1=-\frac{f_m b}{f}\,e^0\wedge e^2\wedge e^3 .
\end{equation}

\paragraph{Second block $I_2=\sum_{A<B}\delta_\ell\Sigma^{AB}\wedge\delta_m\omega_{AB}$.}
The non-vanishing pairs are $(01),(12),(13),(23)$. Using
\eqref{eq:var_l_Sigma}, \eqref{eq:var_m_omega_low}, and
$e^2\wedge e^3\wedge e^0=+e^0\wedge e^2\wedge e^3$:
\begin{align}
  \delta_\ell\Sigma^{01}\wedge\delta_m\omega_{01}
  &=\big(2\gamma\,\Sigma^{01}\big)\wedge\big(-K_m e^0\big)
  =2\gamma K_m\,e^0\wedge e^2\wedge e^3,\\
  \delta_\ell\Sigma^{12}\wedge\delta_m\omega_{12}
  &=(\beta+\gamma)\big(e^0\wedge e^3\big)\wedge\Big(-\tfrac{\rho'f_m}{2\rho\sqrt f}e^2\Big)
  =(\beta+\gamma)\frac{\rho'f_m}{2\rho\sqrt f}\,e^0\wedge e^2\wedge e^3,\\
  \delta_\ell\Sigma^{13}\wedge\delta_m\omega_{13}
  &=\Big[(\beta+\gamma)\Sigma^{13}-2\sqrt f\cos\theta\,d\phi\wedge e^2\Big]
    \wedge\Big(-\tfrac{\rho'f_m}{2\rho\sqrt f}e^3\Big)
  =(\beta+\gamma)\frac{\rho'f_m}{2\rho\sqrt f}\,e^0\wedge e^2\wedge e^3,
\end{align}
 and
\begin{align}
  \delta_\ell\Sigma^{23}\wedge\delta_m\omega_{23}
  &=\big(2\sqrt f\cos\theta\,d\phi\wedge e^1\big)
    \wedge\Big(-\tfrac{\ell f_m}{\rho^2\sqrt f}e^0\Big)
  =-\frac{2\ell f_m\cos\theta}{\rho^2}\,d\phi\wedge e^1\wedge e^0
  =+\frac{2\ell f_m\cos\theta}{\rho^2}\,dt\wedge dr\wedge d\phi .
\end{align}
Here we used $d\phi\wedge e^1\wedge e^0=-dt\wedge dr\wedge d\phi$. Collecting, we get
\begin{equation}
  I_2=\Big[2\gamma K_m+(\beta+\gamma)\frac{\rho'f_m}{\rho\sqrt f}\Big]
  \,e^0\wedge e^2\wedge e^3+\frac{2\ell f_m\cos\theta}{\rho^2}\,dt\wedge dr\wedge d\phi .
\end{equation}

Combining the two blocks, we obtain
\begin{equation}
  I=2B'\,e^0\wedge e^2\wedge e^3-\frac{4\ell f_m\cos\theta}{\rho^2}\,dt\wedge dr\wedge d\phi,
  \label{eq:current}
\end{equation}
with
\begin{equation}
  B'\equiv-\frac{f_m b}{f}-2\gamma K_m-(\beta+\gamma)\frac{\rho'f_m}{\rho\sqrt f}.
  \label{eq:Bprime}
\end{equation}
Substituting \eqref{eq:params} and \eqref{eq:abbrev} into \eqref{eq:Bprime},
the $f_\ell$-dependent pieces cancel and
\begin{equation}
  B'=\frac{2\ell}{\rho^4\sqrt f}.
  \label{eq:Bvalue}
\end{equation}

The presymplectic current must satisfy $dI=0$ on shell. This provides an
independent consistency check of \eqref{eq:current}. Writing
$e^0=\sqrt f\,dt+2\ell\sqrt f\cos\theta\,d\phi$ and
$e^0\wedge e^1\wedge e^2\wedge e^3=\rho^2\sin\theta\,dt\wedge dr\wedge d\theta\wedge d\phi$,
one finds that the only surviving component of $dI$ is
\begin{equation}
  dI=\Big\{
    2\Big[-\sqrt f\,\partial_r B'-B'\Big(\tfrac{f'}{2\sqrt f}+\tfrac{2\rho'\sqrt f}{\rho}\Big)\Big]\rho^2
    +\frac{4\ell f_m}{\rho^2}
  \Big\}\sin\theta\,dt\wedge dr\wedge d\theta\wedge d\phi .
\end{equation}
With \eqref{eq:Bvalue} the bracket evaluates to
\begin{equation}
  2\rho^2\cdot\frac{4\ell\rho'}{\rho^5}-\frac{8\ell r}{\rho^4}
  =\frac{8\ell r}{\rho^4}-\frac{8\ell r}{\rho^4}=0,
\end{equation}
so $dI=0$ identically. The current \eqref{eq:current} is closed.

Using $f_m=-2r/\rho^2$ and $\rho^2=r^2+\ell^2$, \eqref{eq:current} becomes
\begin{equation}
  I=\frac{4\ell\sin\theta}{\rho^2}\,dt\wedge d\theta\wedge d\phi
    +\frac{8\ell r\cos\theta}{\rho^4}\,dt\wedge dr\wedge d\phi .
\end{equation}
This is a total derivative:
\begin{equation}
  I=d\Big[\frac{4\ell}{\rho^2}\cos\theta\,dt\wedge d\phi\Big],
  \label{eq:primitive}
\end{equation}
as follows from $\partial_r(\rho^{-2})=-2r/\rho^4$ and
$\partial_\theta\cos\theta=-\sin\theta$. Therefore
\begin{equation}
  \Omega(\partial_m,\partial_\ell)
  =\frac{1}{16\pi G}\int_\Sigma I
  =\frac{1}{16\pi G}\oint_{\partial\Sigma}\frac{4\ell}{\rho^2}\cos\theta\,dt\wedge d\phi .
\end{equation}
On every constant-$t$ cross-section of the boundary (the sphere at infinity,
and likewise the horizon) the pullback of $dt$ vanishes, so the integrand is
zero. Equivalently, $\oint_{S^2}\cos\theta\,d\phi=0$. Hence
\begin{equation}
 \Omega(\partial_m,\partial_\ell)=0 .
\end{equation}
The mass $m$ and the NUT charge $\ell$ are \emph{not} symplectically conjugate
in the metric phase space.
\end{widetext}


\begin{thebibliography}{99}
\bibitem{Cal65} M. G. Calkin, An invariance property of the free electromagnetic field, Am. J. Phys. {\bf33}, 958 (1965).
\bibitem{DT76} S. Deser and C. Teitelboim,  Duality transformations of Abelian and non-Abelian gauge fields, Phys. Rev. D {\bf13}, 1592 (1976).
\bibitem{Ran92} A. F. Ra\~nada, Topological electromagnetism, J. Phys. A: Math. Gen. {\bf25}, 1621 (1992).
\bibitem{CB12} R. P. Cameron and S. M. Barnett, Electric-magnetic symmetry and Noether's theorem,
New J. Phys. {\bf14}, 123019  (2012).
\bibitem{BBN13} K. Y. Bliokh, A. Y. Beksaev, and F. Nori, Dual electromagnetism: Helicity, spin, momentum and angular momentum, New J. Phys. {\bf15},  033026 (2013).
\bibitem{AS96} G. N. Afanasiev and Yu. P. Stepanovsky, The helicity of the free electromagnetic field and its physical meaning,  Nuovo Cimento A  {\bf109},  271 (1996).
\bibitem{EDHZ16} M. Elbistan, C. Duval, P. A. Horv\'athy, and P.-M. Zhang, Duality and helicity: A symplectic viewpoint, Phys. Lett. B {\bf761}, 265 (2016).
\bibitem{CW87} C. Crnkovic and E. Witten, Covariant description of canonical formalism in geometrical theories, in  {\it Three Hundred Years of Gravitation}, edited by S. W. Hawking and W. Israel,  Cambridge University Press, 1987, pp. 676-684.
\bibitem{Bar14} S. M. Barnett, Maxwellian theory of gravitational waves and their mechanical properties,  New J. Phys. {\bf16}, 023027 (2014).
\bibitem{AAB21} S. Aghapour, L. Andersson, and R. Bhattacharyya, Helicity and spin conservation in linearized gravity, Gen Relativ. Gravit.  {\bf53}, 102 (2021).
\bibitem{Tot22} G. Z. T\'oth, Energy-momentum tensor and duality symmetry of linearized gravity in the Fierz formalism, Class. Quantum Grav. {\bf39}, 075003 (2022).
\bibitem{DS05} S. Deser and D. Seminara, Free spin $2$ duality invariance cannot be extended to general relativity, Phys. Rev. D {\bf71}, 081502 (2005).

\bibitem{Kol22} U. Kol, Duality in Einstein's gravity, arXiv:2205.05752.
\bibitem{NY82} H. T. Nieh and M. L. Yan, An identity in Riemann-Cartan geometry, J. Math. Phys. {\bf23}, 373 (1982).
\bibitem{LZ26} J. Long and X.-H. Zhou, Reduction of topological invariants on null hypersurfaces, JHEP01(2026)116.
\bibitem{Bar95} J. F. Barbero G., Real Ashtekar variables for Lorentzian signature space times, Phys. Rev. D {\bf51}, 5507 (1995).
\bibitem{Imm97} G. Immirzi,  Real and complex connections for canonical gravity, Class. Quantum Grav. {\bf14}, L177 (1997).
\bibitem{Holst} S. Holst, Barbero's Hamiltonian derived from a generalized Hilbert-Palatini action, Phys. Rev. D {\bf53}, 5966 (1996).
\bibitem{Mercuri09} S. Mercuri, Fermions in the Ashtekar--Barbero connection formalism for arbitrary values of the Immirzi parameter, Phys. Rev. D {\bf73}, 084016 (2006).
\bibitem{DKS09}  G. Date, R. K. Kaul, and S. Sengupta, Topological interpretation of Barbero-Immirzi parameter, Phys. Rev. D {\bf79}, 044008 (2009).
\bibitem{RW12} C. Rovelli and E. Wilson-Ewing, Discrete symmetries in covariant loop quantum gravity, Phys. Rev. D {\bf86}, 064002 (2012).
\bibitem{Bianchi24} E. Bianchi and M. Rincon-Ramirez, Spinfoams, $\gamma$-duality and parity violation in primordial gravitational waves, Phys. Rev. D {\bf113}, 124013 (2026).
\bibitem{KolYau23} U. Kol and S.-T. Yau, Duality in gauge theory, gravity and string theory, arXiv:2311.07934.
\bibitem{Monteiro23} R. Monteiro, No $U(1)$ 'electric-magnetic' duality in Einstein gravity, JHEP04(2024)093.
\bibitem{Ash86} A. Ashtekar, New variables for classical and quantum gravity, Phys. Rev. Lett. {\bf57}, 2244 (1986); New Hamiltonian formulation of general relativity, Phys. Rev. D {\bf36}, 1587 (1987).
\bibitem{DH97} B. P. Dolan and K. P. Haugh, A covariant approach to Ashtekar's canonical gravity, Class. Quantum Grav. {\bf14}, 477 (1997).
\bibitem{LSZ98} Y. Luo, M.-X. Shao, and Z.-Y. Zhu, Diffeomorphism invariance of geometric descriptions of Palatini and Ashtekar's gravity, Phys. Lett. B {\bf419}, 37 (1998).
\bibitem{IW} V. Iyer and  R. M. Wald, Some properties of Noether charge and a proposal for dynamical black hole entropy. Phys. Rev. D {\bf50}, 846 (1994).
\bibitem{Griffiths07} J. B. Griffiths and J. Podolsky, A note on the parameters of the Kerr--NUT--(anti-)de Sitter spacetime, Class. Quant. Grav. {\bf24}, 1687, (2007).
\bibitem{Geroch71} R. P. Geroch, A method for generating solutions of Einstein's equations, J. Math. Phys. {\bf12}, 918 (1971).
\end{thebibliography}
\end{document}